\documentclass[twocolumn]{revtex4}[12pt]
\usepackage{graphicx}
 \usepackage{amsmath, amsthm, amssymb, amsfonts}
 \usepackage{ragged2e}
\begin{document}
\justify
\title{Modulational instability and ion-acoustic envelope solitons in four component plasmas}
\author{$^*$N. A. Chowdhury, M. M. Hasan, A. Mannan,  and A. A. Mamun}
\address{Department of Physics, Jahangirnagar University, Savar,
Dhaka-1342, Bangladesh.\\
$^*$Email: nurealam1743phy@gmail.com}
\begin{abstract}
  Modulational instability (MI) of ion-acoustic  waves (IAWs) has been theoretically
  investigated in a plasma system which is composed of inertial warm adiabatic ions, isothermal positrons, and
  two temperature superthermal electrons. A nonlinear Schr\"{o}dinger (NLS) equation is derived
  by using reductive perturbation method that governs the MI of the IAWs.
  The numerical analysis of the solution of NLS equation shows the existence of both
  stable (dark envelope solitons exist) and unstable (bright envelope solitons exist) regimes of IAWs.
  It is observed that the basic features (viz. stability of the wave profile and MI growth rate)
  of the IAWs are significantly modified by the superthermal parameter ($\kappa$)
  and related plasma parameters. The results of our present investigation should be useful
  for understanding different nonlinear phenomena in both space and laboratory plasmas.
\end{abstract}
\maketitle
\section{Introduction}
%%%%%%%%%%%%%%%%%%%%%%%%%%%%%%%%%%%%%%%%%%%%%%%%%%%%%%%%%%%%%%%%%%%%%%%%%%%%%%%%%%%%%%%%%%%%%%%%%%%%%%%%%%%%%%%%%%%%%
During the last few decades, the research about electron-positron-ion (e-p-i) plasma  has been spectacularly
increasing because the  observational (Viking Satellite \cite{Temerin1982} and THEMIS mission \cite{Ergun1998})
evidence exposed the existence of large amount e-p-i plasma in the space (solar atmospheres \cite{Goldreich1969,Hansen1988},
pulsar magnetosphere  \cite{Liang1998,Michel1982}, polar regions of neutron stars \cite{Michel1991}) and
laboratories plasmas \cite{Marklund2006}. Many authors encounter with wave dynamics \cite{Popel1995,Shukla2003,Shukla2004,Kourakis2006,Mamun2010a,Mamun2010b,Ferdousi2014,Rahman2014,Rahman2014a}
(such as electron-acoustic waves (EAWs), positron-acoustic waves (PAWs), and  IAWs) to understand the
physics of collective behaviour in such kind of space and laboratories plasmas.

In case of natural space or in laboratories plasmas (i.e., hot, tenuous, and collisionless) high energy
particles may co-exist \cite{Rehman2016} with isothermal distrubuted particles and their characteristics are deviated from the eminent Maxwellian distribution. Sometimes
this type of particles can be modeled by non-Maxwellian high-energy tail distribution which is known
as generalized Lorentzian (kappa) distribution \cite{Vasyliunas1968,Summers1991,Mace1995,Maksimovic1997,Vi˜nas2005,Hellberg2009}.
The kappa distribution and its relation to the Maxwellian distribution was first introduced by
Vasylinuas \cite{Vasyliunas1968}. This type of distribution may be arisen \cite{Alam2014}, due to the external
forces acting on the natural space plasmas or to wave particle interaction. The Lorentzian or kappa distribution
function \cite{Summers1991} in three dimensional case can be written in the form
\begin{eqnarray}
&&\hspace*{-1cm}F_k (\nu)
=\frac{\Gamma(\kappa+1)}{(\pi\kappa\theta^2)^{3/2}\Gamma(\kappa-1/2)}
(1+\frac{\nu^2}{\kappa\theta^2})^{-(\kappa+1)},\label{k}
\end{eqnarray}
where $\theta = [(2\kappa-3)/\kappa]V_t$, is the effective thermal speed  which depends on the
usual thermal velocity $V_t = (k_BT/m)^{1/2}$ , $\Gamma$ is the standard gamma
function, $T$ is the characteristic kinetic temperature, which is the temperature
of the equivalent Maxwellian with the same average kinetic energy \cite{Hellberg2009},
and $k_B$ is the Boltzmann constant. The
parameter $\kappa$ is the measurement of the slope of the energy
spectrum of the superthermal particles forming the tail of the
velocity distribution function which is also called spectral
index. Lower values of $\kappa$ represents a hard spectrum with a
strong non-Maxwellian tail \cite{Summers1991}. The Lorentzian or
kappa distribution function $F_k (\nu)$ reduces to the Maxwellian
(thermal equilibrium distribution) for the limit of large
spectral index \cite {Kaladze2014}, i.e., $\kappa
\rightarrow\infty$ and $\theta\rightarrow V_t = (k_BT/m)^{1/2}$.

A number of works \cite{Baluku2008,Baluku2010,Choi2011,Sultana2011,Shahmansouri2013a} have
been done by considering single-temperature superthermal (kappa
distributed) electrons. But in many space  as well as in laboratory plasmas, electrons are
found to have two distinct temperatures \cite{Goswami1976,Buti1980,Nishihara1981}. Solar wind around 1 AU (Earth's orbit),
high intensity laser irradiation \cite{Estabrook1978},
turbulent of thermonuclear interest, hot cathode discharge \cite{Goswami1976} plasmas are composed of two-electron
populations. By taking two temperature superthermal electrons, Panwar \textit{et al.} \cite{Panwar2014} studied the
oblique ion-acoustic (IA) cnoidal waves in a magnetized plasma. In case of Saturn's magnetosphere, by considering two
temperature kappa distributed electrons IA solitons are studied
by Baluku and Hellberg \cite{Baluku2012}. They found that solitons of both polarities can exist over restricted
ranges of fractional hot electron density ratio. By considering two temperature electron model, Baboolal \textit{et al.} \cite{Baboolal1990}
numerically shown that how exist domains for arbitrary amplitude IA solitons and double layers are determined by cut off
conditions. Shahmansouri and Alinejad \cite{Shahmansouri2013b} studied
the linear and nonlinear excitation of arbitrary amplitude IA solitary waves in a
magnetized plasma comprising of two temperature electrons. They found
that the electron superthermality reduces the phase velocities of both modes. Masud \textit{et al.} \cite{Masud2013}
studied the nonplanar geometry of dust-ion-acoustic solitary waves
containing two populations of thermal electrons in dusty plasma and found that electrons with
different temperatures can significantly modify the wave dynamics. Rehman and Mishra \cite{Rehman2016} analyzed IA solitary waves in e-p-i plasma with
two temperature electrons and isothermal positrons, they found that the ratio of cold to hot electron temperature
plays a crucial role to generate and controlling the shape of solitons.

The investigations of the MI of
IAWs both theoretically and experimentally have been
increasing day by day due to their successful applications in
space  as well as laboratory plasmas.  The
NLS equation have been used to understand different nonlinear phenomena such as single pulse \cite{Sultana2011}
and envelope structures respectively, observed in space and laboratory plasmas \cite{Bailung1998,Nakamura1999,Nakamura2001,Ma2006}.
Recently, a number of authors \cite{Salahuddin2002,Esfandyari2006a,Esfandyari2006b,Kourakis2006,Jehan2008,Gill2010} investigated the
MI and envelope solitons structure in pair and e-p-i plasmas. By using reductive perturbation method (RPM), most of them has obtained envelope
solitons  \cite{Kourakis2006,Esfandyari2006a,Gill2010}. The electrostatic envelope
solitons have also been studied by using Krylov- Bogoliubov-Mitropolsky (KBM) method \cite{Salahuddin2002,Jehan2008} in plasmas.
In unmagnetized electron-ion  plasmas Ju-Kui \textit{et al.} \cite{Ju-Kui2002} has used  RPM,
whereas Durrani \textit{et al.} \cite{Durrani1979} has used KBM method  to study the MI of IAWs
with warm ions. The aim of the present paper is, by using RPM a NLS
equation is derived for nonlinear electrostatic IA waves in unmagnetized
e-p-i plasmas in the presence of warm ions, superthermal electrons with two distinct
temperatures and isothermal positron.

The manuscript is organized as follows: The basic
governing equations of our considered plasma model is presented in Sec. II. By using reductive perturbation technique,
we derive a NLS equation which governs the slow amplitude evolution
in space and time is given in Sec. III.  The stability analysis is presented, in Sec. IV. The envelope solitons
are presented in sec. V. The discussion is  provided in Sec. VI.
%%%%%%%%%%%%%%%%%%%%%%%%%%%%%%%%%%%%%%%%%%%%%%%%%%%%%%%%%%%%%%%%%%
\section{Governing Equations}
We consider an unmagnetized plasma system comprising of inertial warm adiabatic ions, isothermal
positrons, and two temperature superthermal electrons (hot and cold). At equilibrium, the quasi-neutrality condition
can be expressed as $n_{i0} + n_{p0}=  n_{h0}+n_{c0}$, where $n_{i0}$, $n_{p0}$, $n_{h0}$ and $n_{c0}$ are
the equilibrium number densities of warm adiabatic ion, isothermal  positron, and superthermal  hot electron
and cold electron, respectively. The normalized equations governing the IAWs in our considered plasma system are given by
\begin{eqnarray}
&&\hspace*{-2cm}\frac{\partial n_i}{\partial t}+\frac{\partial}{\partial x}(n_i u_i)=0,\label{a1}\\
&&\hspace*{-2cm}\frac{\partial u_i}{\partial t} + u_i\frac{\partial u_i }{\partial x}=-\frac{\partial \phi}{\partial x}-3 \alpha n_i \frac{\partial n_i}{\partial x},\label{a2}\\
&&\hspace*{-2cm}\frac{\partial^2 \phi}{\partial x^2}=n_c+\gamma n_h-\sigma n_i-(1-\sigma+\gamma) n_p. \label{a3}
\end{eqnarray}
For inertialess cold electron, we can obtain the expressions for cold electron  number densities as
 \begin{eqnarray}
&&\hspace*{-1.3cm}n_c=\left[1-\frac{\phi}{(\kappa -3/2)}\right]^{-\kappa+{1/2}}\nonumber\\
&&\hspace*{-.85cm}=1+ C_1 \phi+ C_2 \phi^2 + C_3 \phi^3+\cdot\cdot\cdot\cdot, \label{Ch1}
\end{eqnarray}
where
\begin{eqnarray}
&&\hspace*{-1.5cm}C_1=\left( \frac{\kappa-{1/2}}{\kappa-{3/2}}\right),~~C_2=\frac{1}{2}\left(\frac{\kappa^2-{1/4}}{(\kappa-{3/2})^2}\right),\nonumber\\
&&\hspace*{-1.5cm}C_3=\frac{1}{6}\left(\frac{(\kappa-{1/2})(\kappa+{1/2})(\kappa+{3/2})}{(\kappa-{3/2})^3}\right). \nonumber\
\end{eqnarray}
For inertialess hot electron, we can obtain the expressions for hot electron number densities as
\begin{eqnarray}
&&\hspace*{-1.1cm}n_h= \left [1-\frac{\mu \phi}{(\kappa -3/2)}\right]^{-\kappa+{1/2}}\nonumber\\
&&\hspace*{-.6cm}=1+  C_1 \mu \phi+  C_2 \mu^2 \phi^2 +  C_3 \mu^3 \phi^3 +\cdot\cdot\cdot\cdot.\label{Ch4}
\end{eqnarray}
Similarly for inertialess isothermal positron, we can obtain the expressions for positron number densities as
\begin{eqnarray}
&&\hspace*{-2.0cm} n_p= \mbox{exp}~(-\lambda \phi)\nonumber\\
&&\hspace*{-1.15cm} =1 - \lambda \phi+\frac{ \lambda^2 \phi^2}{2}-\frac{ \lambda^3 \phi^3}{6}+\cdot\cdot\cdot\cdot\label{Ch4}\label{Ch4}
\end{eqnarray}
Substituting equations $(5)-(7)$ into equation $(4)$, and expanding up to third order, we get
\begin{eqnarray}
&&\hspace*{-0.4cm}\frac{\partial^2 \phi}{\partial x^2}=1+\gamma-\eta-\sigma n_i+\gamma_1 \phi+\gamma_2 \phi^2+\gamma_3 \phi^3+\cdot\cdot, \label{Ch4}
\end{eqnarray}
where
\begin{eqnarray}
&&\hspace*{-1.2cm}\eta=(1-\sigma+\gamma),~~~\gamma_1=\gamma \mu C_1+ C_1+\eta \lambda, \nonumber\\
&&\hspace*{-1.2cm}\gamma_2=\gamma \mu^2 C_2+ C_2-(\eta \lambda^2)/2, \nonumber\\
&&\hspace*{-1.2cm}\gamma_3=\gamma \mu^3 C_3+ C_3+(\eta \lambda^3)/6, \nonumber\
\end{eqnarray}
and

~~~~$\alpha=\frac{T_i}{T_c}$,~~$\gamma=\frac{n_{h0}}{n_{c0}}$,~~$\lambda=\frac{T_c}{T_p}$~~$\sigma=\frac{n_{i0}}{n_{c0}}$,~~$\mu=\frac{T_c}{T_h}$.

\noindent In the above equations, $n_i$ is the ion number density normalized by its equilibrium value $n_{i0};$
$u_i$ is the ion fluid speed normalized by the IA wave speed $C_i=(T_c/m_i)^{1/2}$ (with $m_i$ being
the ion rest mass). $T_c, T_h, T_p $ and $T_i$ corresponds to the
temperature of cold electrons, hot electrons, isothermal positrons and ions, respectively. $\phi$ is
the electrostatic wave potential normalized by $T_c/e$ (with $e$ being the magnitude of single
electron charge). The time and space variables are normalized by ${\omega^{-1}_{pi}}=(m_i/4\pi e^2 n_{c0})^{1/2}$
and $\lambda_{Di}=(T_c/4 \pi e^2 n_{c0})^{1/2}$, respectively.
%%%%%%%%%%%%%%%%%%%%%%%%%%%%%%%%%%%%%%%%%%%%%%%%%%%%%%%%%%%%%%%%%%%%%%%%%%%%%%%%%%%%%%%%

%%%%%%%%%%%%%%%%%%%%%%%%%%%%%%%%%%%%%%%%%%%%%%%%%%%%%%%%%%%%%%%%%%%%%%%%%%%%%%%%%%%%%%%%%%%%%%%%
\section{Derivation of the NLS Equation}
To study the modulation of the IAWs in our considered plasma system, we will derive the NLS equation by
employing the RPM. So, we first introduce
the independent variables are stretched as
\begin{eqnarray}
&&\hspace*{-3.9cm}\xi={\epsilon}(x  - v_g t),~~~\tau={\epsilon}^2 t, \label{eq6}
\end{eqnarray}
where $v_g$ is the envelope group velocity  and $\epsilon~(0<\epsilon<1)$ is a small (real) parameter.
Then we can write a general expression for the dependent variables \cite{Elwaki2010} as
\begin{eqnarray}
&&\hspace*{-1.0cm}G(x,t)=G_0 +\sum_{m=1}^{\infty}\epsilon^{(m)}\sum_{l=-\infty}^{\infty}G_{l}^{(m)}(\xi,\tau) ~\mbox{exp}(i l\Theta), \nonumber\\
&&\hspace*{-1.0cm}G_l^{(m)}=[n_{il}^{(m)}, u_{il}^{(m)}, \phi_l^{(m)}]^T,~~G_l^{(0)}=[1, 0, 0]^T, \label{eq11}
\end{eqnarray}
where  $\Theta=(kx-\omega t)$, simultaneously k and $\omega$ are real variables representing the carrier wave number
and frequency, respectively.  $ G_l^{(m)}$ satisfies the pragmatic condition $ G_l^{(m)}= G_{-l}^{(m)^*}$, where the asterisk denotes
the complex conjugate. The derivative operators in the above equations are treated as follows:
\begin{eqnarray}
&&\hspace*{-1.6cm}\frac{\partial}{\partial t}\rightarrow\frac{\partial}{\partial t}-\epsilon v_g \frac{\partial}{\partial\xi}+\epsilon^2\frac{\partial}{\partial\tau},~~\frac{\partial}{\partial x}\rightarrow\frac{\partial}{\partial x}+\epsilon\frac{\partial}{\partial\xi}. \label{eq10}
\end{eqnarray}
Substituting equations $(9)-(11)$ into equations $(2),(3)$, and $(8)$ and the first order $(m=1)$ equations with $(l=1)$, gives
\begin{eqnarray}
&&\hspace*{-3.3cm}-i\omega n_1^{(1)}+iku_1^{(1)}=0,\nonumber\\
&&\hspace*{-3.3cm}-i\omega u_1^{(1)}+ik\phi_1^{(1)}+ik\Omega n_1^{(1)}=0,\nonumber\\
&&\hspace*{-3.3cm}\sigma n_1^{(1)}-k^2\phi_1^{(1)}-\gamma_1 \phi_1^{(1)}=0,\label{eq14}
\end{eqnarray}
where $\Omega=3\alpha$. The solution for the first harmonics read as
\begin{eqnarray}
&&\hspace*{-2.4cm} n_1^{(1)}=\frac{k^2}{S}\phi_1^{(1)},~~~~~~~u_1^{(1)}=\frac{k \omega}{S}\phi_1^{(1)}, \label{eq11}
\end{eqnarray}
where $S=\omega^2-k^2 \Omega$. We thus obtain the dispersion relation for IAWs
\begin{eqnarray}
&&\hspace*{-4.2cm} \omega^2=\frac{\sigma k^2}{(k^2+\gamma_1)}+k^2\Omega.\label{eq12}
\end{eqnarray}
The second-order when $(m=2)$ reduced equations with $(l=1)$ are
\begin{eqnarray}
&&\hspace*{-1.0cm}n_1^{(2)}=\frac{k^2}{S}\phi_1^{(2)}+\frac{2ik\omega(v_g k-\omega)}{S^2} \frac{\partial \phi_1^{(1)}}{\partial\xi},\nonumber\\
&&\hspace*{-1.0cm}u_1^{(2)}=\frac{k \omega}{ S}\phi_1^{(2)} +\frac{i(v_g k-\omega)(\omega^2 +k^2 \Omega)}{S^2} \frac{\partial \phi_1^{(1)}}{\partial\xi}, \label{eq16}
\end{eqnarray}
with the compatibility condition
\begin{eqnarray}
&&\hspace*{-4.4cm}v_g=\frac{\partial \omega}{\partial k}=\frac{\sigma \omega^2-S^2}{k\sigma\omega}.\label{eq17}
\end{eqnarray}
The amplitude of the second-order harmonics are found to be proportional to $|\phi_1^{(1)}|^2$
\begin{eqnarray}
&&\hspace*{-2.1cm}n_2^{(2)}=C_4|\phi_1^{(1)}|^2,~~~~~~~n_0^{(2)}=C_7 |\phi_1^{(1)}|^2,\nonumber\\
&&\hspace*{-2.1cm}u_2^{(2)}=C_5 |\phi_1^{(1)}|^2,~~~~~~~u_0^{(2)}=C_8|\phi_1^{(1)}|^2,\nonumber\\
&&\hspace*{-2.1cm}\phi_2^{(2)}=C_6 |\phi_1^{(1)}|^2,~~~~~~~\phi_0^{(2)}=C_9 |\phi_1^{(1)}|^2,\label{eq19}
\end{eqnarray}
where
\begin{eqnarray}
&&\hspace*{-2mm}C_4=\frac{\Omega k^6  +3\omega^2 k^4 +2C_6 S^2 k^2 }{2S^3}, \nonumber\\
&&\hspace*{-2mm}C_5=\frac{\omega C_4 S^2  -\omega k^4 }{k S^2},  \nonumber\\
&&\hspace*{-2mm}C_6=\frac{\sigma(\Omega k^6 +3\omega^2 k^4)-2\gamma_2 S^3}{2S^3{(4k^2+\gamma_1)-2\sigma S^2 k^2}}, \nonumber\\
&&\hspace*{-2mm}C_7=\frac{2\omega v_g k^3 +\Omega k^4+ \omega^2 k^2+C_9 S^2}{S^2(v^2_g-\Omega)}, \nonumber\\
&&\hspace*{-2mm}C_8=\frac{ v_g C_7 S^2-2\omega k^3 }{S^2},  \nonumber\\
&&\hspace*{-2mm}C_9=\frac{2\sigma \omega v_g k^3+\Omega \sigma k^4 +\sigma k^2\omega^2 -2\gamma_2 S^2(v^2_g-\Omega)}{\gamma_1 S^2(v^2_g-\Omega)-\sigma S^2}.  \nonumber\
\end{eqnarray}
Finally, the third harmonic modes $(m=3)$ and $(l=1)$ and  with the help of  equations $(13) - (17)$,
give a system of equations, which can be reduced to the following  NLS equation:
\begin{eqnarray}
&&\hspace*{-3.4cm}i\frac{\partial \Phi}{\partial \tau}+P\frac{\partial^2 \Phi}{\partial \xi^2}+Q|\Phi|^2\Phi=0, \label{eq24}
\end{eqnarray}
where $\Phi=\phi_1^{(1)}$ for simplicity. The dispersion coefficient $P$ is
\begin{eqnarray}
&&\hspace*{-0.5cm}P=\frac{v_g\Omega^2 k^5 -3 v_g k \omega^4+4\Omega k^2 \omega^3 +2v_g\Omega \omega^2  k^3  -4\omega \Omega^2 k^4 }{2\sigma \omega^2 k^2},\nonumber\
\end{eqnarray}
and the nonlinear coefficient $Q$ is
%\begin{widetext}
%\[
%Q=\frac{S}{2\sigma \omega k^2}\left[2S \gamma_2(C_6+C_9)+3S \gamma_3-\frac{(\sigma \omega^2 k^2+\sigma \Omega k^4 )(C_4+C_7)}{S}-\frac{2\sigma \omega k^3(C_5+C_8)}{S}\right].\linebreak\\
%\]
%\end{widetext}
\begin{eqnarray}
&&\hspace*{-0.7cm} Q=\frac{S}{2\sigma \omega k^2}\left[-\frac{(\sigma \omega^2 k^2+\sigma \Omega k^4 )(C_4+C_7)}{S} \right.\nonumber\\
&&\hspace*{0.3cm}\left.+2S \gamma_2(C_6+C_9)+3S \gamma_3-\frac{2\sigma \omega k^3(C_5+C_8)}{S}\right].\nonumber\
\end{eqnarray}
%%%%%%%%%%%%%%%%%%%%%%%%%%%%%%%%%%%%%%%%%%%%%%%%%%%%%%%%%%%%%%%%%%%%%%%%%%%%%%%%%%%%%%%%%%%%%%%%%%%%%%%%%%%%%%%%%%%%%%%%%%%%%%%
\begin{figure*}[htp]
  \centering
  \begin{tabular}{ccc}
  \includegraphics[width=78mm]{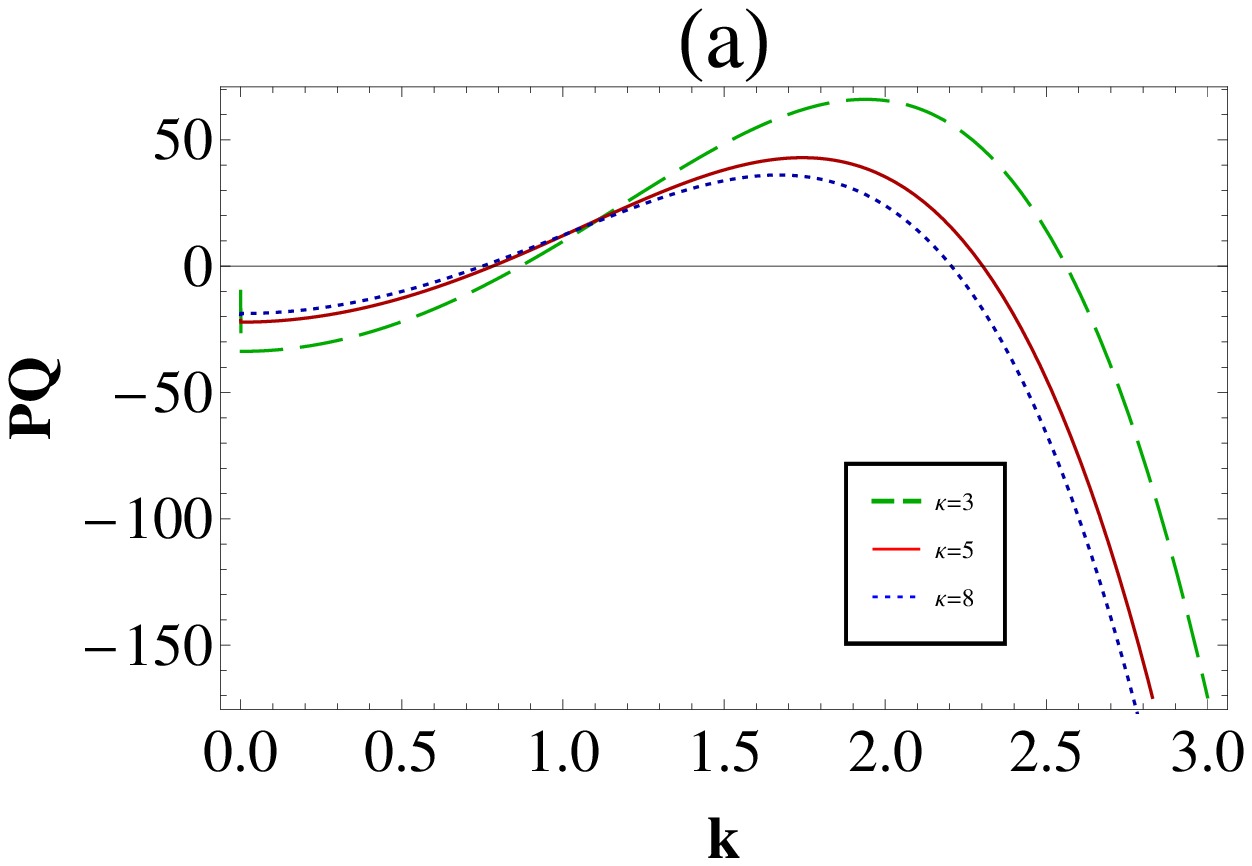}&
  \hspace{0.15in}
  \includegraphics[width=80mm]{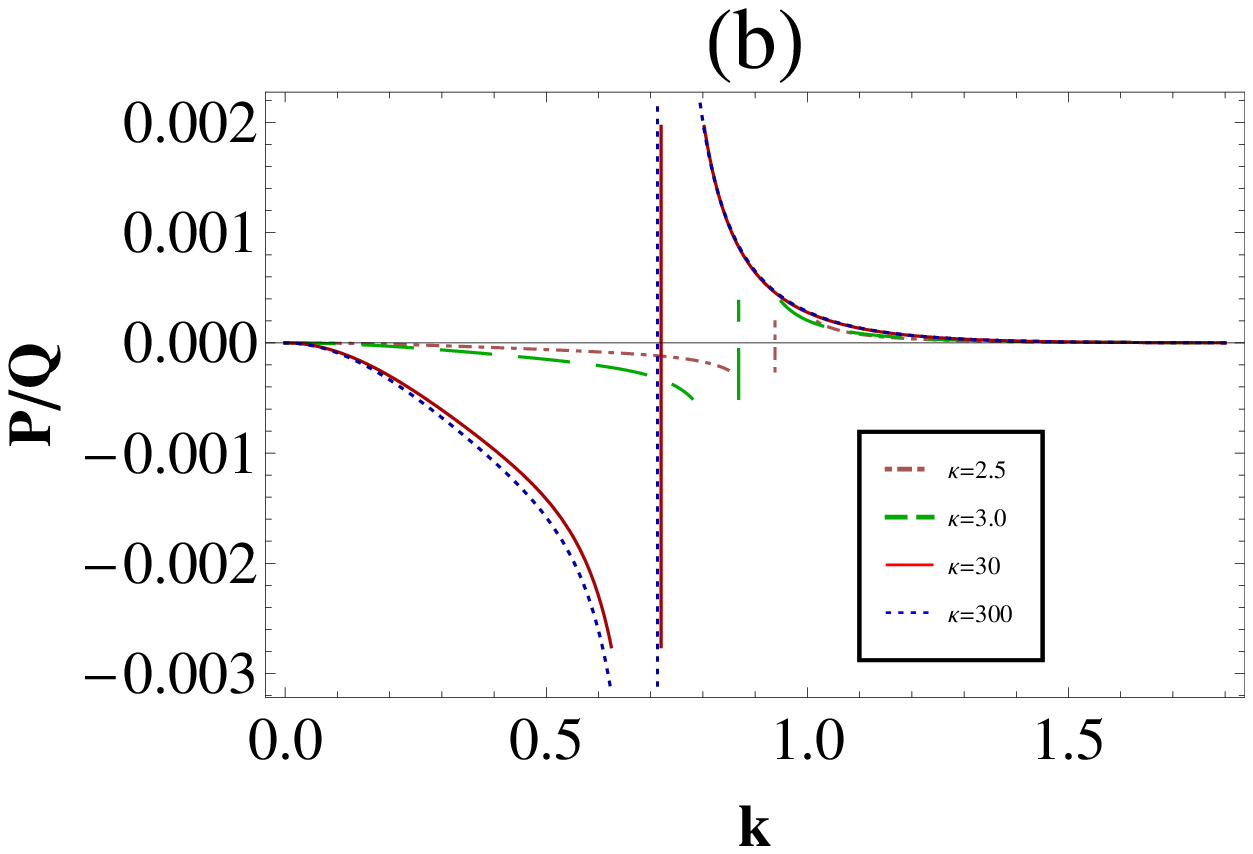}\\
  \end{tabular}
  \label{figur}\caption{(a) Showing the variation of $PQ$ against $k$ for different values of $\kappa$.
  (b) Plot of $P/Q$ against $k$ for different values of $\kappa$.
  All the figures are generated by using these values, $\alpha=0.11, \gamma=0.85, \lambda=0.1, \sigma=0.3,\mu=0.11$, and $\kappa=3$.}
\end{figure*}
\begin{figure*}[htp]
  \centering
  \begin{tabular}{ccc}
  \includegraphics[width=80mm]{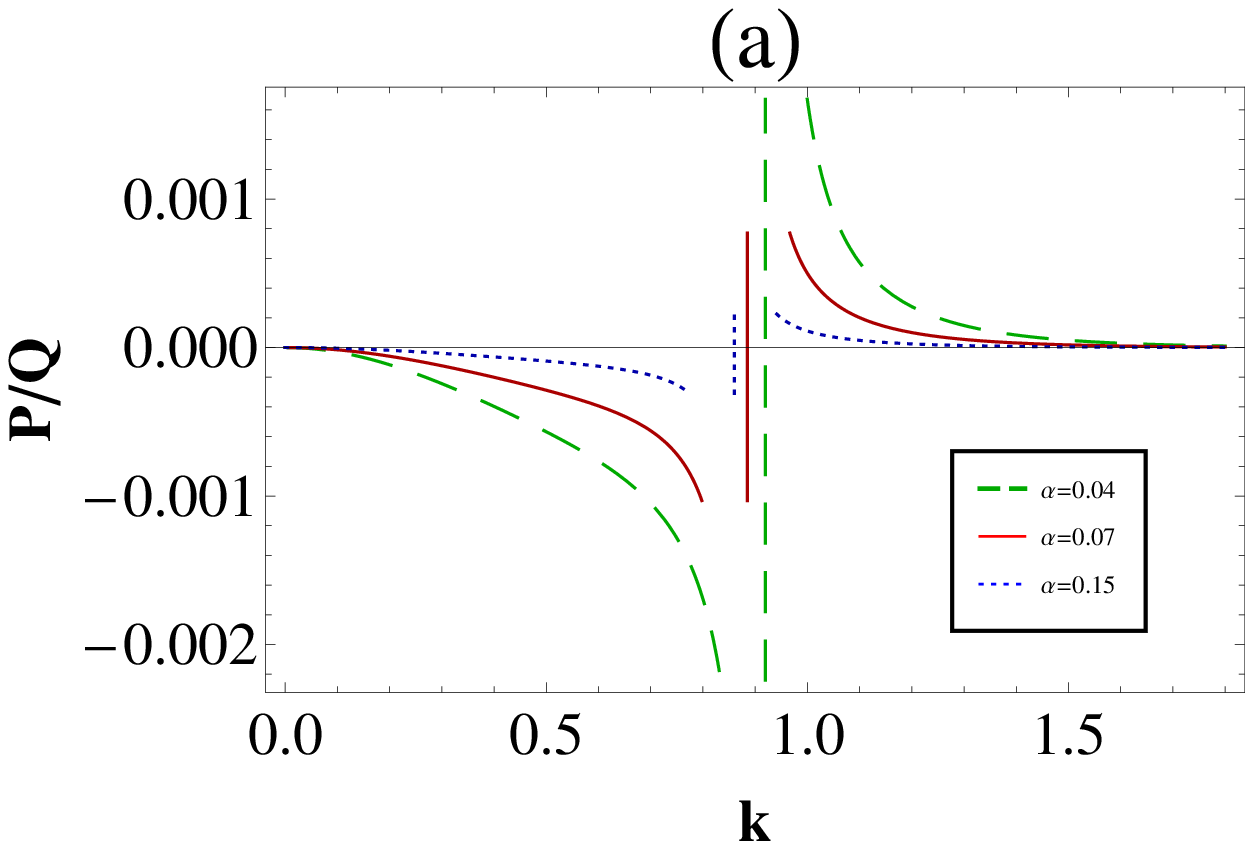}&
  \hspace{0.15in}
  \includegraphics[width=80mm]{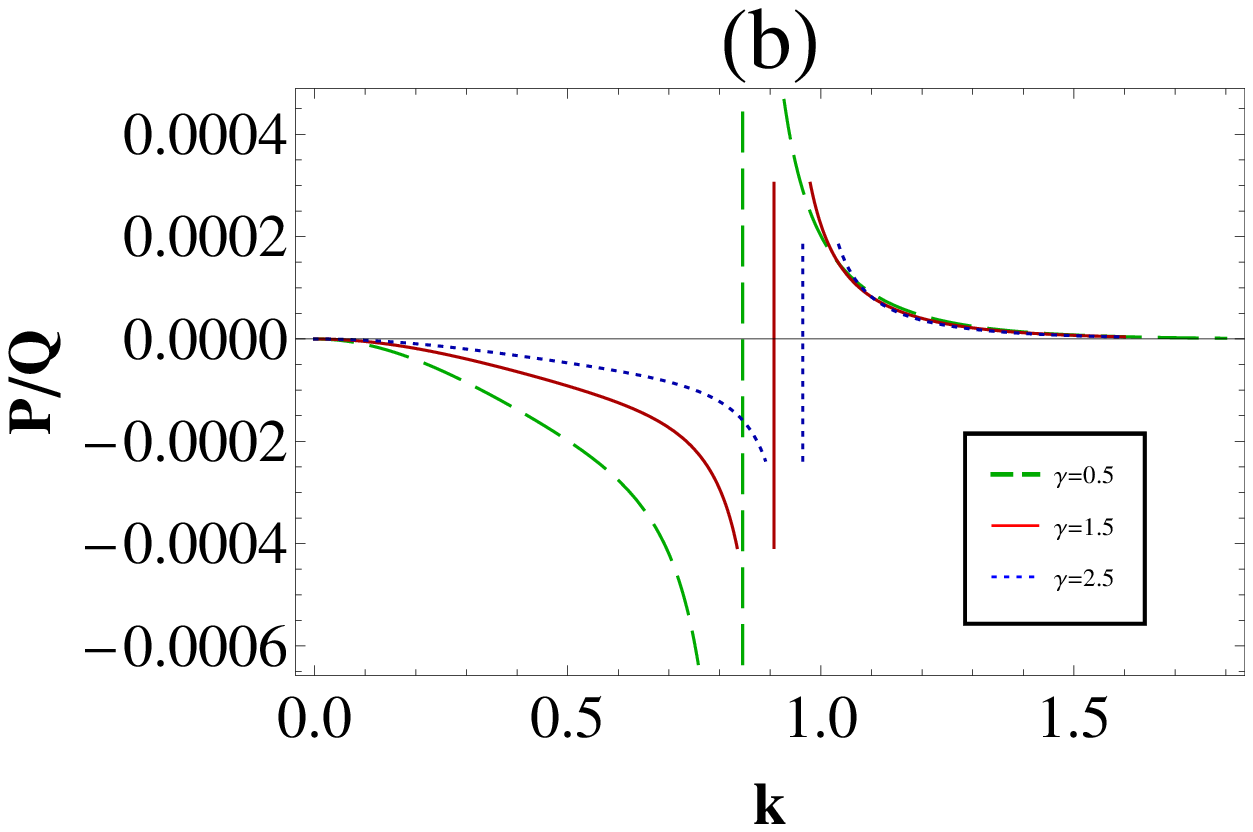}\\
  \end{tabular}
  \label{figur}\caption{Showing the variation of $P/Q$ against $k$ for different values of plasma parameters,
  (a) For $\alpha$, (b) For $\gamma$.}
\end{figure*}
\begin{figure*}[htp]
  \centering
  \begin{tabular}{ccc}
  \includegraphics[width=83mm]{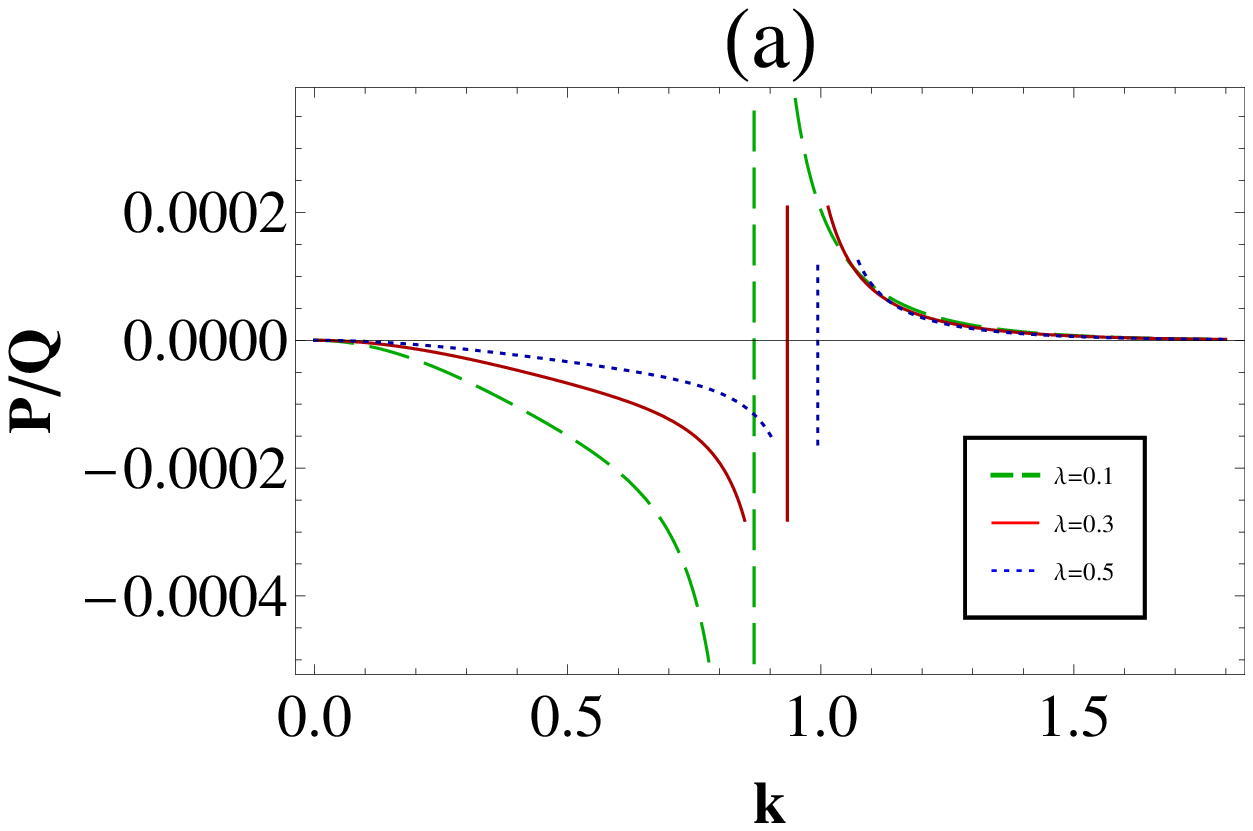}&
  \hspace{0.15in}
  \includegraphics[width=80mm]{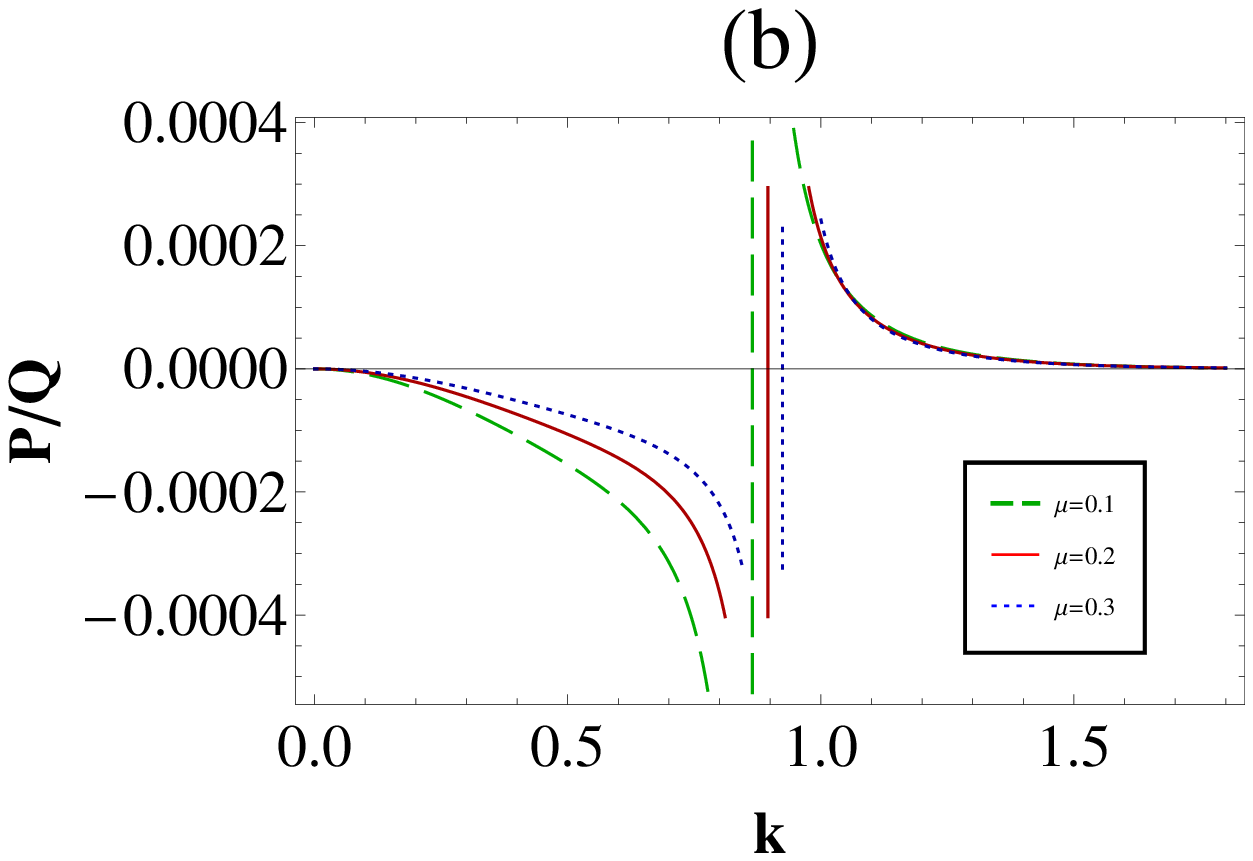}\\

  \includegraphics[width=80mm]{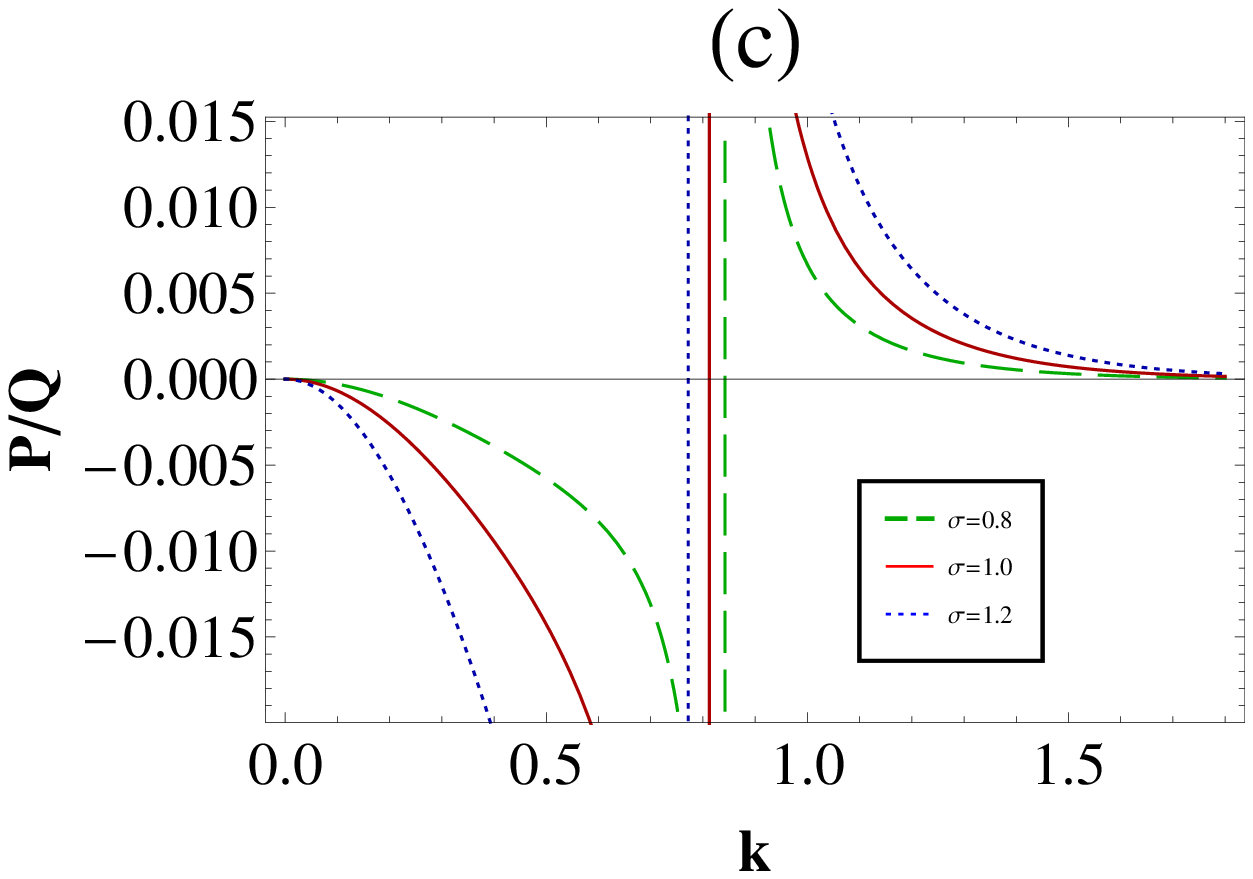}&
  \hspace{0.15in}
  \includegraphics[width=76mm]{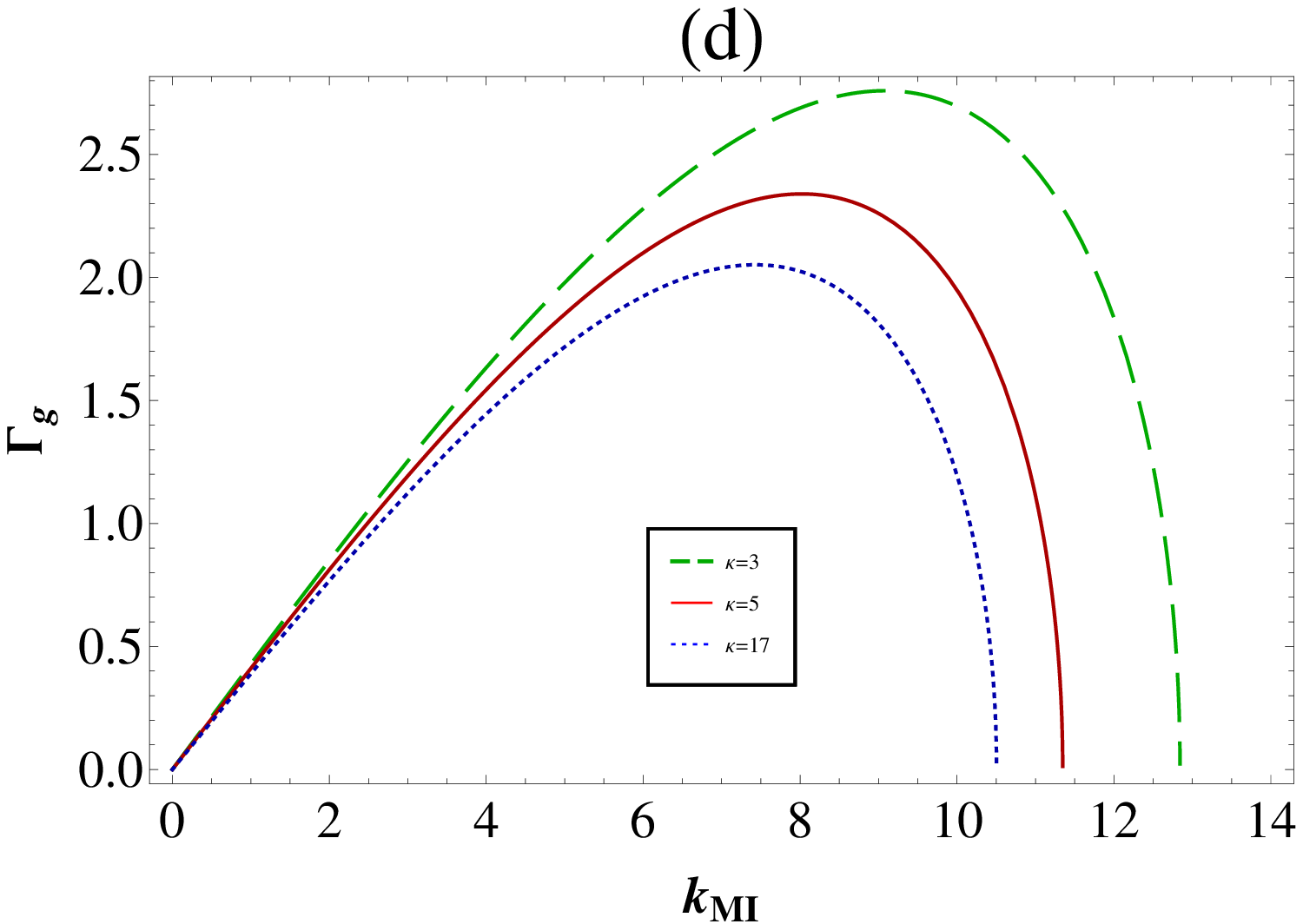}\\
  \end{tabular}
  \label{figur}\caption{Showing the variation of $P/Q$ against $k$ for different plasma parameters,
  (a) For $\lambda$, (b) For $\mu$, and (c) For $\sigma$. (d) Plot of the of MI growth rate $(\Gamma_g)$
  against ${k_{MI}}$ for different values of $\kappa$. Along with $k=1.2$ and $\Phi=0.06$.}
\end{figure*}
\begin{figure*}[htp]
  \centering
  \begin{tabular}{ccc}
  \includegraphics[width=80mm]{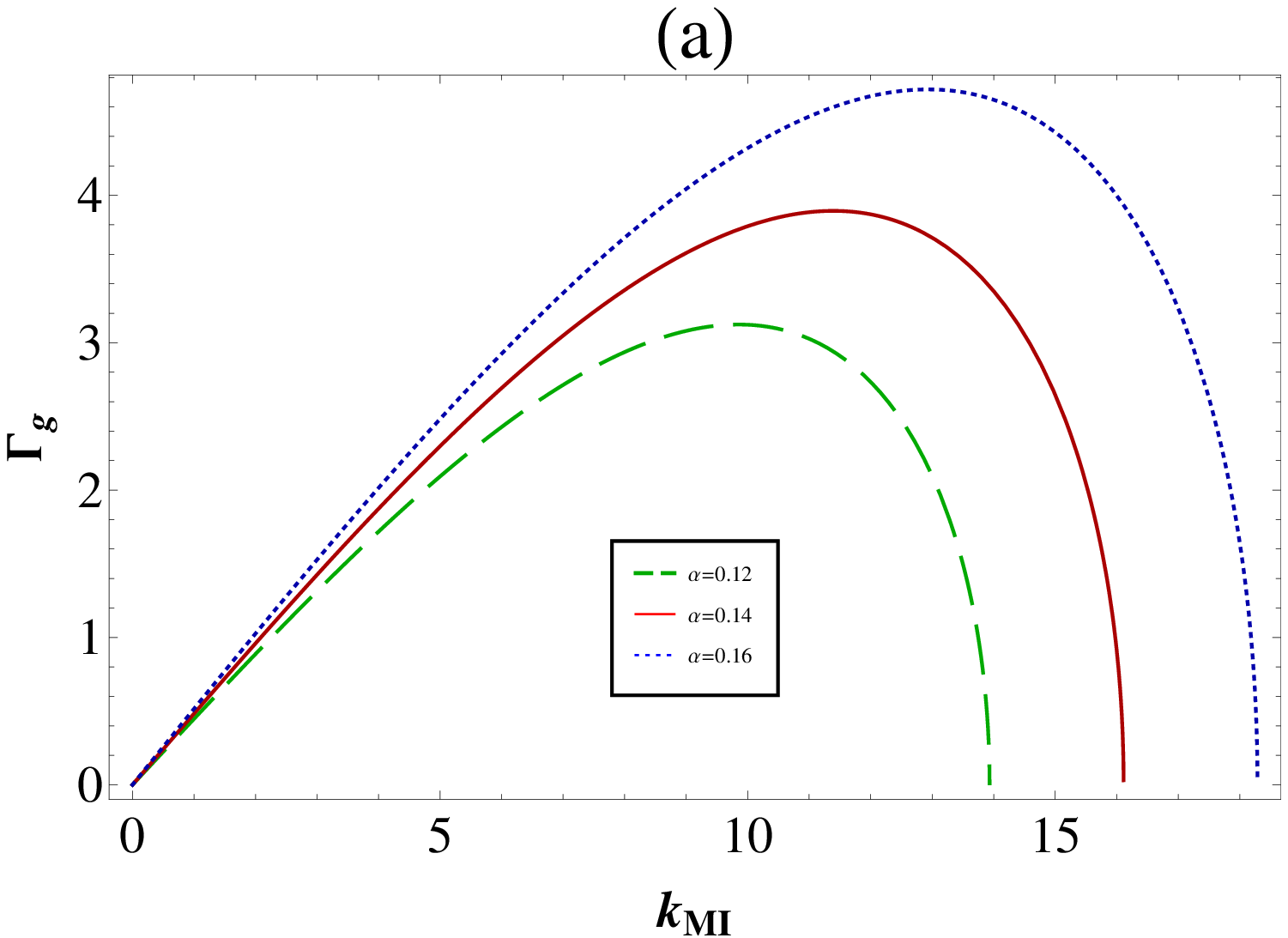}&
  \hspace{0.15in}
  \includegraphics[width=80mm]{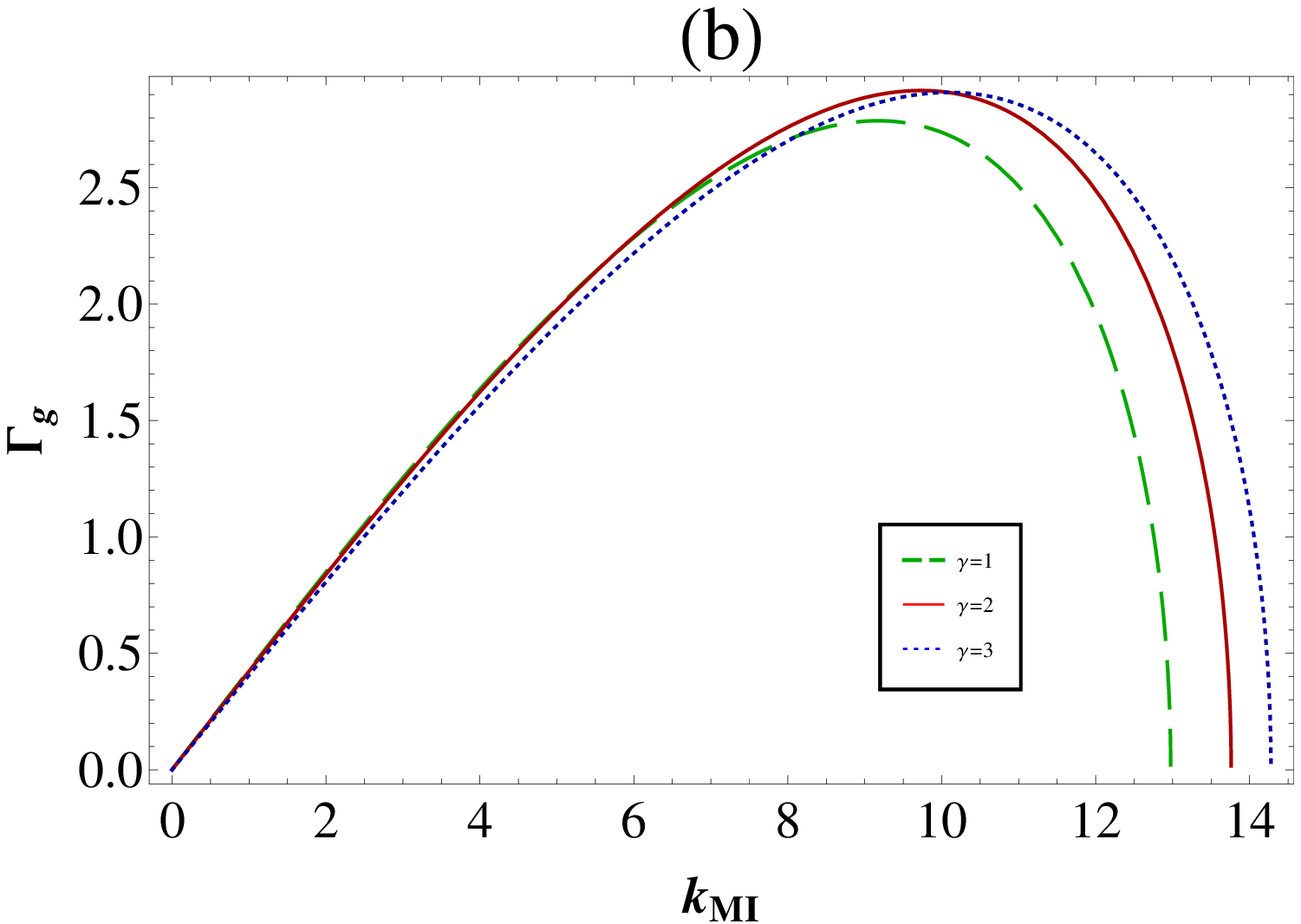}\\
  \end{tabular}
  \label{figur}\caption{Plot of the of MI growth rate $(\Gamma_g)$ against ${k_{MI}}$ for different plasma parameter,
  (a) For $\alpha=3$, (b) For $\gamma$. Along with $k=1.2$ and $\Phi=0.06$.}
\end{figure*}
\begin{figure*}[htp]
  \centering
  \begin{tabular}{ccc}
  \includegraphics[width=80mm]{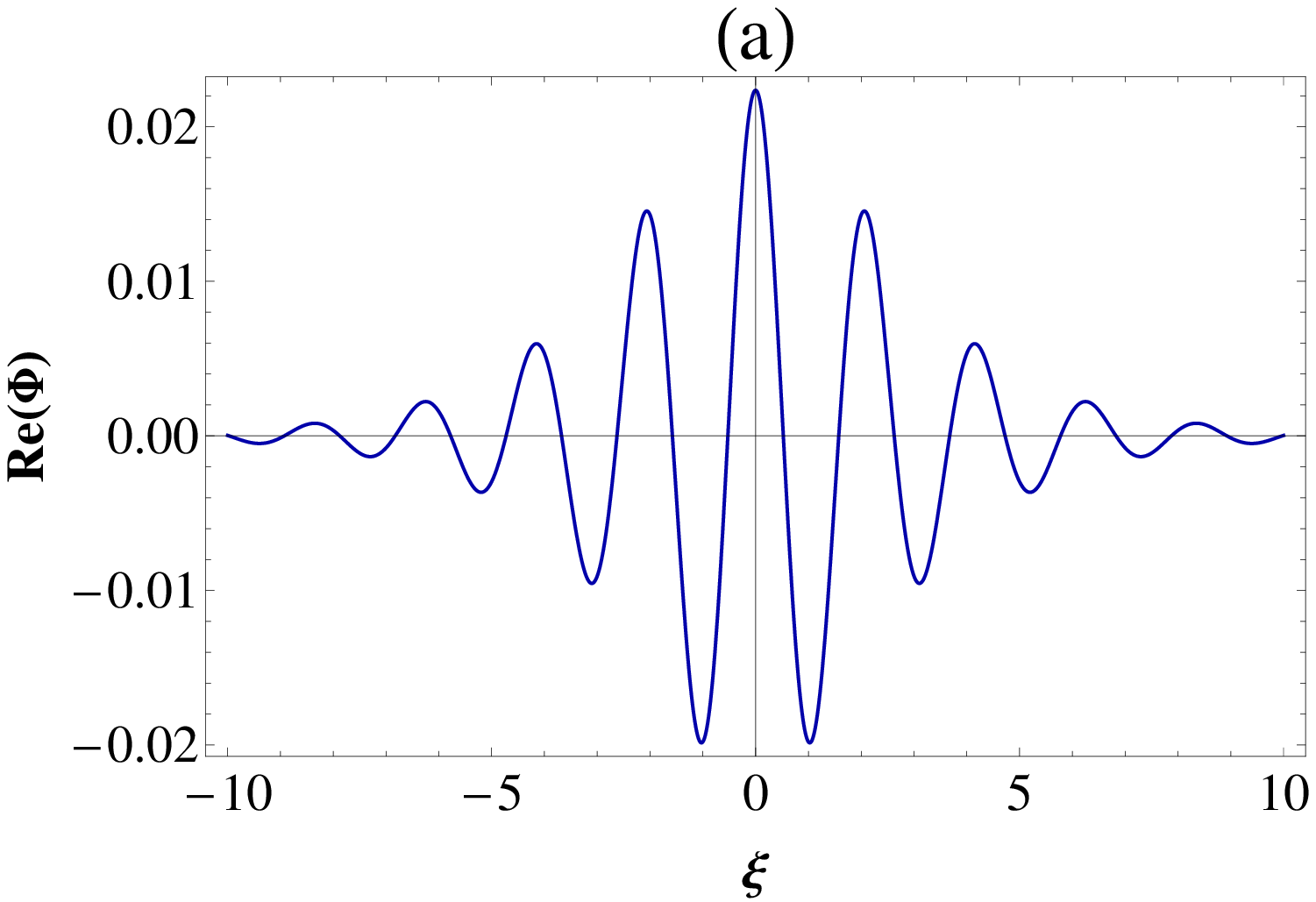}&
  \hspace{0.15in}
  \includegraphics[width=80mm]{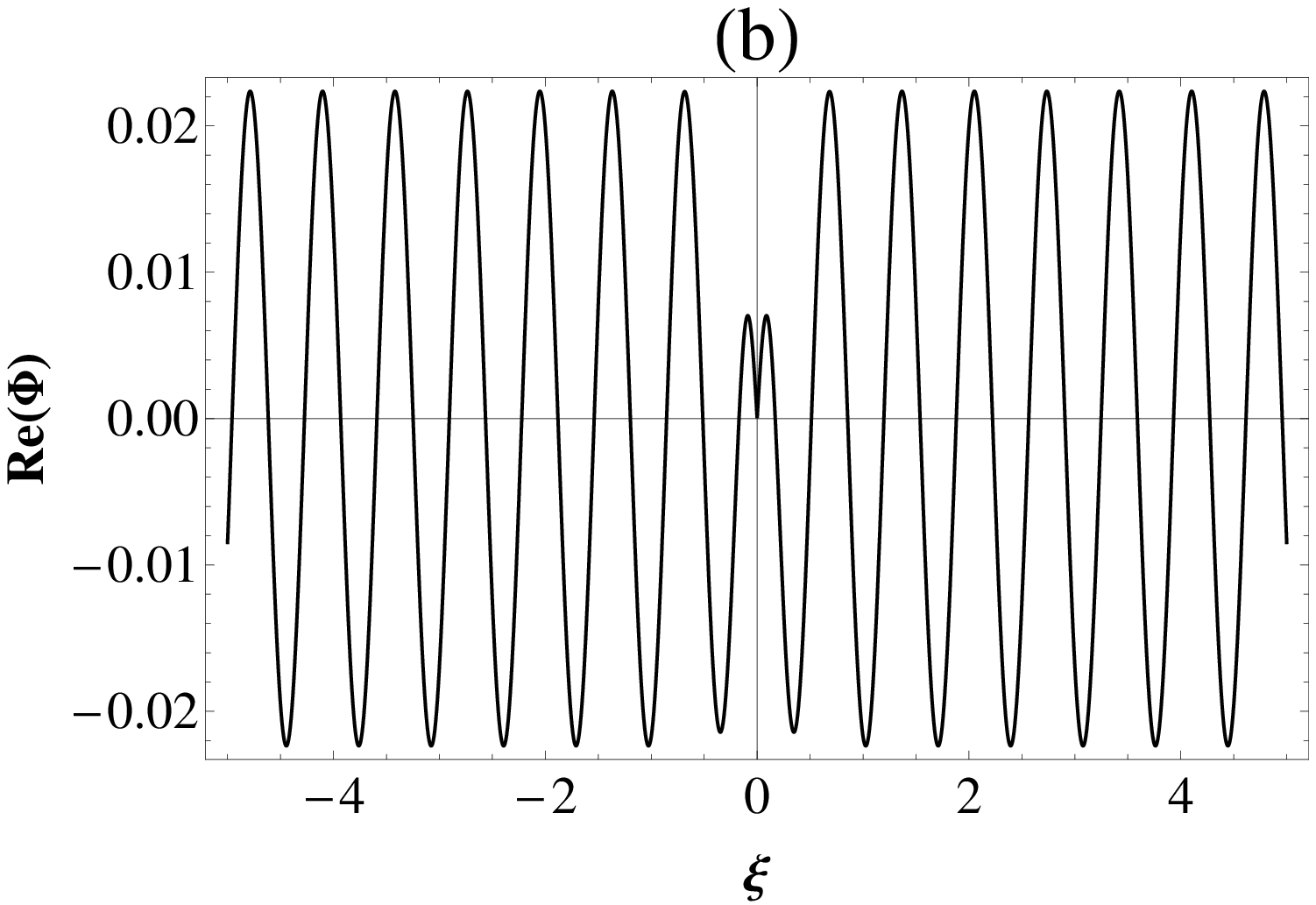}\\
  \end{tabular}
  \label{figur}\caption{ Envelope solitons solution of the NLS equation $(18)$.
  (a) Bright envelope solitons for $k=0.9$, (b) Dark envelope solitons for $k=0.1$,
  along with $\psi_{0}=0.0005, U=0.3, \tau=0,$ and $\Omega_0=0.4$.}
\end{figure*}
%%%%%%%%%%%%%%%%%%%%%%%%%%%%%%%%%%%%%%%%%%%%%%%%%%%%%%%%%%%%%%%%%%%%%%%%%%%%%%%%%%%%%%%%%%%%%%%%%%%%%%%%%%%%%%%%%%%%%%%
\section{Stability analysis}
The evolution of IAWs is governed by the equation $(18)$, essentially depends on the coefficients product $PQ$.
Let us consider the harmonic modulated amplitude solution $\Phi=\Phi_o \exp(iQ{|\Phi_o|^2}\tau)$.
Following the standard stability analysis, one may perturb the amplitude by setting
$\Phi=\hat{\Phi}_0+\epsilon \hat{\Phi}_{1,0} \exp [i({k_{MI}}\xi-{\omega_{MI}}\tau)]+c.c$ (the
perturbation wave number ${k_{MI}}$ and the frequency ${\omega_{MI}}$ should be distinguished
from their carrier wave homolog quantities, denoted by $k$ and $\omega$). Hence, the nonlinear
dispersion relation for the amplitude modulation \cite{Sultana2011,Sabry2008,Schamel2002} is
\begin{eqnarray}
&&\hspace*{-2.1cm}{\omega^2_{MI}}=P^2{k^2_{MI}} \left ({k^2_{MI}}-2\frac{Q}{P}{|\Phi_o|^2}\right).\label{eq12}
\end{eqnarray}
Clearly, if $PQ<0$, ${\omega_{MI}}$ is always real for all values of ${k_{MI}}$, hence in this region the
IAWs is stable in the presence of small perturbation. On the other hand, when $PQ>0$ , the MI
would set in as ${\omega_{MI}}$ becomes imaginary and the envelope is unstable for
${k_{MI}}<k_c=\sqrt{2Q{|\Phi_o|}^2/P}$, where $k_c$ is the critical value of the wave
number of modulation and $\Phi_o$ is the amplitude of the carrier waves. In the region $PQ>0$ and ${k_{MI}}<k_c$, the
growth rate ($\Gamma_g$) of  MI is given by
\begin{eqnarray}
&&\hspace*{-3.5cm}\Gamma_g=|P|~{k^2_{MI}}\sqrt{\frac{k^2_{c}}{k^2_{MI}}-1}.\label{eq112}
\end{eqnarray}
Clearly, the maximum value $\Gamma_{g(max)}$ of $\Gamma_g$ is obtained at ${k_{MI}}=k_c/\sqrt{2}$
and is given by $\Gamma_{g(max)}=|Q||\Phi_0|^2$.

The coeffficients of dispersion term $P$ and nonlinear term $Q$ are dependent on  various plasma parameters,
such as $\alpha, \beta, \gamma, \sigma, \mu, \lambda $ and $\kappa $. Thus, these parameters may be controlled
the stability conditions of the IAWs. Therefore, we have investigated
the stability of the  profile by depicting the ratio of $P/Q$ versus $k$ for different plasma parameters.
When the sign of the ratio $P/Q$ is negative, the modulated envelope pulse is stable, while the sign of
the ratio $P/Q$ is positive, the modulated envelope pulse will be unstable against external perturbations.
It is clear that both stable and unstable region are obtained from the figures $2-4$.
When $P/Q\rightarrow\pm\infty$, the corresponding value of $k(=k_c)$ is called critical or threshold
wave number for the onset of MI. This critical value separates the  unstable ($P/Q>0$) from the stable region ($P/Q<0$) one.
\section{Envelope solitons}
If $PQ<0$,  the modulated envelope pulse is stable and in this region dark envelope
solitons exist, on the other hand when $PQ>0$, the modulated envelope pulse which is unstable against external
perturbations and lead to formation of  bright envelope solitons. A solution of  equation $(18)$
may be sought in the form $\Phi=\sqrt{\psi}~\mbox{exp}(i\theta)$, where $\psi$ and $\theta$ are real variables
which are determined by substituting into the NLS equation and separating real and imaginary parts. An
interested reader is referred to  \cite{Sultana2011,Schamel2002,Kourakis2005,Sukla2002, Fedele2002,Shalini2015} for details.
The different types of solution thus obtained are clearly summarized in the following paragraphs.
\subsection{Bright solitons}
When $PQ>0$, we find bright envelope solitons. The general analytical form of bright solitons reads
\begin{eqnarray}
&&\hspace*{-2.8cm}\psi=\psi_{0}~ \mbox{sech}^2 \left(\frac{\xi-U\tau}{W}\right),\nonumber\\
&&\hspace*{-2.8cm}\theta=\frac{1}{2P} \left[U \xi+ \left(\Omega_0-\frac{U^2}{2}\right)\tau\right].\label{eq112}
\end{eqnarray}
Here, $U$ is the propagation speed (a constant), $W$ is the soliton width, and $\Omega_0$ oscillating frequency for $U=0$.
Figure $6$(a) represents the bright envelope solitons.
\subsection{Dark solitons}
When $PQ<0$, we find dark envelope solitons whose general analytical form  reads as
\begin{eqnarray}
&&\hspace*{-2.1cm}\psi=\psi_{0}~ \mbox{tanh}^2 \left(\frac{\xi-U\tau}{W}\right),\nonumber\\
&&\hspace*{-2.1cm}\theta=\frac{1}{2P}\left[ U \xi-\left( \frac{U^2}{2}-2PQ\psi_{0}\right)\tau \right].
\end{eqnarray}
Interestingly, in both of the latter two equation, the relation between soliton width  $W$ and the
constant maximum amplitude $\psi_{0}$ are related by
\begin{eqnarray}
&&\hspace*{-5.1cm}W=\sqrt{\frac{2|P/Q|}{ \psi_{0}}}.
\end{eqnarray}
\noindent The ratio $P/Q$ determines the soliton width $W$ as $\psi_0 W \sim (P/Q)^{1/2}$. So lower $P/Q$ values suggest narrower solitons and vice versa.
Figure $6$(b) represents the dark envelope solitons.
\section{Discussion }
In this work, we have considered an unmagnetized four-component plasma consisting of inertial warm adiabatic ions, isothermal
positrons, and two temperature superthermal electrons (hot and cold). By employing the reductive perturbation method,
a NLS equation is derived, which governs the evolution of IAWs. We have investigated the existence of both stable and unstable
regions for IAWs structures and the associated MI of electrostatic wave packets. The results, we have found from this
 investigation which can be summarized as follows:
\begin{enumerate}
\item{The variation of $PQ$ with $k$ for different values of superthermality (via $\kappa$)
  is depicted in  Fig. $1(a)$. One can recognize that when $P$ and $Q$ are opposite sign ($PQ<0$),
  there is a  stable region (the IAWs are modulationally stable) whereas $P$ and $Q$ are same sign ($PQ>0$), there
  is an  unstable region (the IAWs are modulationally unstable). With the increasing of the values of $\kappa$ the unstable region is decreasing.
  The intersecting point of the $PQ$ curve with the $k$-axis is called critical or threshold wave number $(k_c)$.}
\item{The $k_c$ value is greatly controlled by superthermality (via $\kappa$). It may be noted that the smaller value of $\kappa$ means strong superthermality.
  With the increase of $\kappa$, the value of $k_c$   is decreased, which is depicted in  Fig. $1(b)$.
  For a large value of $\kappa=30$ or $100$, the $k_c=0.71$ remains almost constant. But if $\kappa <3$, the value of $k_c$  is changed
  rapidly. So stability of the wave profile is so much sensitive to change with $\kappa$, when $\kappa\leq3$.}
\item{ The effects of ion temperature (via $\alpha$) on the wave profile is extremely high to change the stability of the electrostatic wave packets.
  It is observed from  Fig. $2(a)$ that with the  increasing of ion temperature the $k_c$ is shifted to the lower value that means excited ions
  minimize the stability region for IAWs. So ion temperature plays a crucial role for controlling the
  stability of the IAWs profile.}
\item{In  Fig. $2(b)$ the variation of $P/Q$ with $k$ has been plotted for different values of hot electron concentration (via $\gamma$).
  We see that $k_c$ increases with the increasing of hot electron concentration, the critical value is shifted to higher value. That means higher concentration of hot
  electron provides greater restoring force which extend the stable region.}
\item{It can be observed from the  Fig. $3(a)$, the stability of the IAWs profile is also governed by the positron temperature
  (via $\lambda$) of our considered plasma model. If the positron temperature of the system  increases,
  then the  value of $k_c$ also decreases. For small wave number there is dark envelope solitons exists whereas bright
  envelope solitons exists for large wave number.}
\item{The effects of the cold electron temperature (via $\mu$) on the stability of IAWs profile is analyzed from Fig. $3(b)$,
  which depicts the dependence of ratio $P/Q$ on $k$ for different values of $\mu$. As cold electron temperature increases the $k_c$ value  is increased. }
\item{The dependence of ratio $P/Q$ on $k$ for different values of ion number density (via $\sigma$) is depicted in  Fig. $3(c)$. Ion number density plays
  an important role to control the stability of the profile. Excess number of ion cause to provide large moment of inertia that may be suppressed the stability region.}
\item{It is observed from Fig. $3(d)$ that MI growth rate are significantly effected by the values of superthermality (via $\kappa$).
  With increasing superthermality, the MI growth rates appear to decrease. The lower values of $\kappa$ (excess superthermality) may be enhanced the MI growth rate.}
\item{The dependence of the MI on ion temperature (via $\alpha$)  is shown in Fig. $4(a)$. With the increase of
  ion temperature, the growth rate of the instability increases. From Fig. $4(b)$, similar behaviour (the maximum value
  of the growth rate increases, with the increasing of hot electron number density) is also observed (via $\gamma$).
  So $\alpha$  and $\gamma$ are enhanced the instability. Moreover, the growth rate $(\Gamma_g)$ increases with
  increasing of ${k_{MI}}$. For a particular value of ${k_{MI}}$, the growth rate  $(\Gamma_g)$ is reached it's
  critical value $(\Gamma_g\equiv\Gamma_{gc})$. Hence the growth rate $(\Gamma_g)$ sharply decreases with further increases the values of ${k_{MI}}$.}
\end{enumerate}
A large number of observations clearly reveal the existence of high-energy/superthermal electrons in various natural space
environment (Saturn's magnetosphere, magnetotail, auroral zones, the ionosphere, solar wind, strong radiation in the interstellar
or interplanetary medium etc.) and laboratories plasmas. We are optimistic that our  nonlinear analysis will be helped to understand
the nonlinear structures (bright and dark envelope solitons) that may be formed in both space and laboratory plasmas which containing of isothermal
positrons, two distinct temperature superthermal electrons (hot and cold), and inertial warm adiabatic ions.
\section*{Acknowledgement}
N. A. Chowdhury is grateful to the Bangladesh Ministry of Science and Technology for
awarding the National Science and Technology (NST) Fellowship.
%%%%%%%%%%%%%%%%%%%%%%%%%%%%%%%%%%%%%%%%%%%%%%%%%%%%%%%%%%%%%%%%%%%%%%%%%%%%%%%%%%%%%%%%%

\end{document}